\documentclass[aps, prl, twocolumn, amsmath,amssymb]{revtex4-2}

\usepackage{graphicx}
\usepackage{hyperref}
\usepackage{dcolumn}
\usepackage{bm}

\hypersetup{
    colorlinks=true, 
    linkcolor=blue,  
    citecolor=blue, 
    urlcolor=blue 
}

\begin{document}

\preprint{APS/123-QED}

\title{Connecting Shear Thinning and Dynamic Heterogeneity in \\ Supercooled Liquids by Localized Elasticity}

\author{Ke-Qi Zeng}
\author{Dong-Xu Yu}
\author{Zhe Wang}
\email{zwang2017@mail.tsinghua.edu.cn}
\affiliation{Department of Engineering Physics and Key Laboratory of Particle and Radiation Imaging (Tsinghua University) of Ministry of Education, Tsinghua University, Beijing 100084, China}

\date{\today}

\begin{abstract}
Supercooled liquids exhibit complicated dynamical behaviors: 
At the microscopic level, the dynamics is heterogeneous spatially, known as dynamic heterogeneity. 
At the macroscopic level, the shear viscosity $\eta$ decreases as shear rate $\dot{\gamma}$ increases with a power law $\eta\sim\dot{\gamma}^{-\lambda}$, known as shear thinning. 
The relation between these two universal dynamical phenomena remains elusive. 
With simulations of several model liquids in two and three dimensions, we show that they are quantitatively bridged by localized elasticity embodied as transient clusters that elastically respond to shear. 
Prominent dynamic heterogeneity emerges right after the massive yielding of these clusters, which is initiated by shear transformation zones and facilitated by elasticity-mediated interaction. 
With this picture, a scaling law relating shear thinning to the characteristic length of dynamic heterogeneity is found.
\end{abstract}

\maketitle

Supercooled liquids exhibit remarkable shear thinning when subjected to strong steady shear flow: 
At high enough shear rates $\dot{\gamma}$, the shear viscosity $\eta$ decreases with $\dot{\gamma}$ as  $\eta\sim\dot{\gamma}^{-\lambda}$ \cite{larson1, oswald1}. 
The microscopic mechanism of this nonlinearity has attracted great scientific interest in the past half century 
\cite{hoover1, hess1, fuchs1, fuchs2, miya1, furu1, furu2, mizuno1, tanaka1, yamaguchi1, yamamoto1, yamamoto2, tanaka2, lubchenko1}. 
Many approaches seek structural indicators by characterizing the structural distortion at the pair level \cite{hess1, fuchs1, fuchs2, miya1, furu1, furu2, tanaka1, mizuno1, yamaguchi1}. 
These efforts connect shear thinning to cage configuration \cite{fuchs1, fuchs2, miya1, furu1, furu2} or 
the distortion of the pair distribution function $g(\bm{r})$ at the spatial range beyond the cage \cite{tanaka1, yamaguchi1}. 
The second class of approaches tackles this problem by considering the dynamic heterogeneity (DH) \cite{yamamoto1, yamamoto2, tanaka2, lubchenko1} 
i.e., dynamic regions where particles collectively undergo nontrivial displacements to rearrange structure \cite{ediger1, royall1}. 
These studies have established a firm correspondence between shear thinning and the evolution of DH \cite{yamamoto1}. 
Such success is expected if one considers the importance of DH in the research on the slow dynamics of supercooled liquids. 
Nevertheless, the relation between the aforementioned two classes of approaches is unclear. 
More seriously, the micro-mechanical connection between shear thinning and dynamic regions remains elusive, which impedes constitutive modeling through DH.

Owing to the extremely viscous nature, deeply supercooled liquids can be viewed as ``solids that flow'' \cite{dyre1, dyre2}, 
in the sense that they behave elastically below certain length scales at short time scales. 
This conceptual picture has been substantiated significantly in recent years \cite{dyre3}. 
For example, elastic features are observed in the stress correlation in supercooled liquids \cite{lemaitre1, fuchs3, egami1, tanaka3}. 
Further studies suggest that the elasticity-mediated interaction plays a key role in facilitating new rearrangements, which finally leads to the emergence of DH \cite{liu1, biroli1, barrat1}. 
Moreover, it is also suggested that the extended-ranged elasticity is crucial in determining the nonlinear rheology of supercooled liquids \cite{egami2, lemaitre2, schw1, cates1}. 
These results open the possibility of connecting shear thinning and DH by elasticity. 

Inspired by some pioneering works \cite{dyre1, egami2}, we recently proposed a concept of localized elastic region (LER) to model the shear thinning of supercooled liquids \cite{wang1}. 
An LER is a transient zone containing hundreds of particles. 
It deforms elastically before yielding, which provides the resistance to the imposed shear \cite{cates1, lacks1, patinet1}. 
In this Communication, we visualize this concept by characterizing the organization of cage jumps, i.e., large nonaffine displacements of particles \cite{biroli2, biroli3}. 
We find that the evolution of the localized elasticity resembles the response of amorphous solids to the start-up shear \cite{oswald1, pete1, egelhaaf1}: 
It sequentially undergoes elastic deformation and plastic deformation, and eventually relaxes due to the avalanche of cage jumps facilitated by the elasticity-mediated interaction. 
This relaxation process gives rise to dynamic regions that compose the major signature of DH. Based on these observations, 
a scaling law connecting the shear thinning and the length scale of DH is given.

To study the dynamics and rheology of supercooled liquids, we employ both Brownian dynamics (BD) simulation and molecular dynamics (MD) simulation, 
which respectively represent colloidal systems and atomic systems. 
All studied systems are binary mixtures. 
The BD system is the same as the one we employed in our previous study \cite{wang1}. 
Its potential parameters well describe the inter-particle interaction of a three-dimensional (3D) charged colloidal suspension \cite{wang2}. 
Two volume fractions of particles, $\phi=45\%$ and $42.5\%$, are adopted. The MD system is the celebrated Kob-Andensen liquid \cite{kob1} at a temperature close to its mode-coupling-theory temperature. 
In addition to the above two 3D systems, we also simulate a two-dimensional (2D) MD system, 
which has been used to study the shear thinning of 2D supercooled liquids \cite{tanaka2}, to test the relevant scaling law. 
The information of the simulation systems, as well as the units of time $\tau_{0}$, length $d_{0}$, stress $\sigma_{0}$ and viscosity $\eta_{0}$ used in following discussion, is given in Methods. 
For all conditions, the SLLOD equation combined with the Lees-Edwards boundary condition \cite{edwards1} is applied for the shear with the stream velocity along 
the $x$ direction and the velocity gradient along the $y$ direction. 
In following parts, we will first establish the connection between DH and shear thinning based on BD results. 
MD results will be appended next, which demonstrate that the conclusions found from 3D BD results also hold well in the atomic system and in 2D. 

\section{Results}

\subsection{DH and convective cluster in flow}

\begin{figure}
    \includegraphics[width=\linewidth]{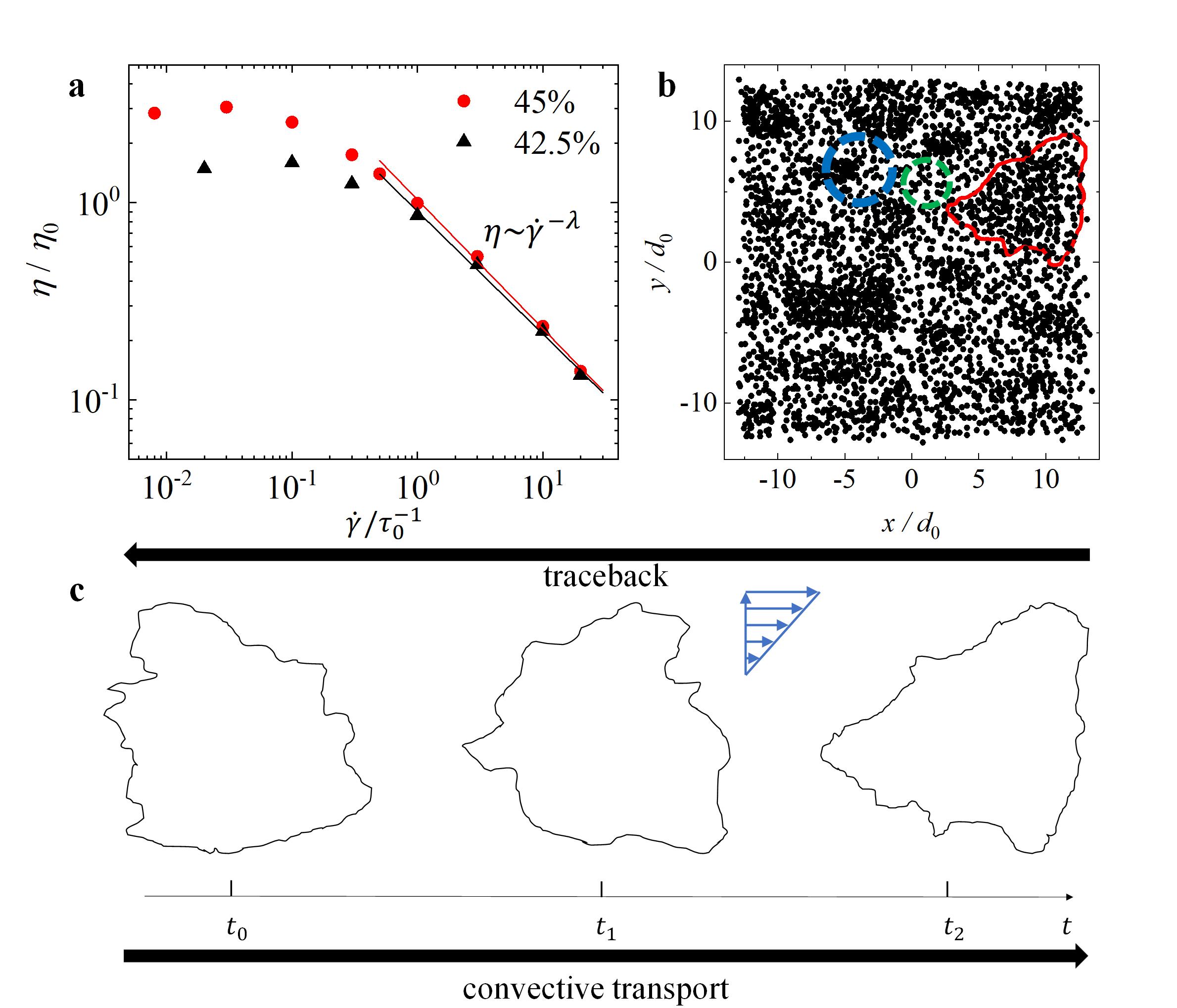}
    \caption{\label{fig:1} 
    \textbf{\textbar Shear thinning, dynamic heterogeneity and convective cluster.} 
    (\textbf{a}) Shear viscosity contributed by the Yukawa interaction for $\phi=42.5\%$ and $45\%$ BD samples. 
    Solid lines denote fits with $\dot{\gamma}^{-\lambda}$ in the shear-thinning regime. 
    (\textbf{b}) Projection of cage-jump events in the shear plane accumulated within a time interval $t_{\chi}$ for the $\phi=45\%$ sample at $\dot{\gamma}\tau_{0}=3$. 
    The red solid line denotes the boundary of a cage-jump cluster (i.e., dynamic region). 
    Moreover, two spherical regions, whose boundaries are respectively denoted by dashed green line and blue line, are marked. 
    The former is outside of any cage-jump cluster. 
    The latter contains the parts of different cage-jump clusters. 
    (\textbf{c}) Convective transport (time-evolution) and traceback (time-reversal) of the boundary highlighted by red solid curve in \textbf{b}.}
\end{figure}

Figure~\ref{fig:1} a shows the shear thinning of the viscosity $\eta$ for two BD samples calculated by:
\begin{equation}
	\eta=-\langle\sum_{i}r_{i,x}f_{i,y} \rangle \big/ V\dot{\gamma} \label{eq1}
\end{equation}
where $\bm{r}_{i}$ and $\bm{f}_{i}$ denote the position of particle $i$ and the deterministic force exerted on particle $i$, respectively, 
and $V$ is the system volume. 
At $\dot{\gamma}\tau_{0}\geq1$, a power-law thinning $\eta\sim\dot{\gamma}^{-\lambda}$ is present. 
To explore the DH in flow, we look for the irreversible nonaffine displacements of particles by identifying cage jumps following Candelier et al. \cite{biroli2, biroli3}. 
Figure~\ref{fig:1}b shows a projection of cage-jump events accumulated within a time interval $t_{\chi}$ for the $\phi=45\%$ BD sample at $\dot{\gamma}\tau_{0}=3$, 
where $t_{\chi}$ is the time at which the dynamic susceptibility $\chi_{4}(t)$ reaches maximum \cite{glotzer1, xu1}. 
The clustering of cage jumps is clearly seen, implying the significance of DH in flow. 
The boundary of such dynamic region is delineated according to the density distribution of cage jumps, and an example is given in Fig.~\ref{fig:1}b (denoted by the red solid line). 
The details for identifying cage jumps and demarcating the boundaries of dynamic regions are given in Methods.

\begin{figure}
    \includegraphics[width=\linewidth]{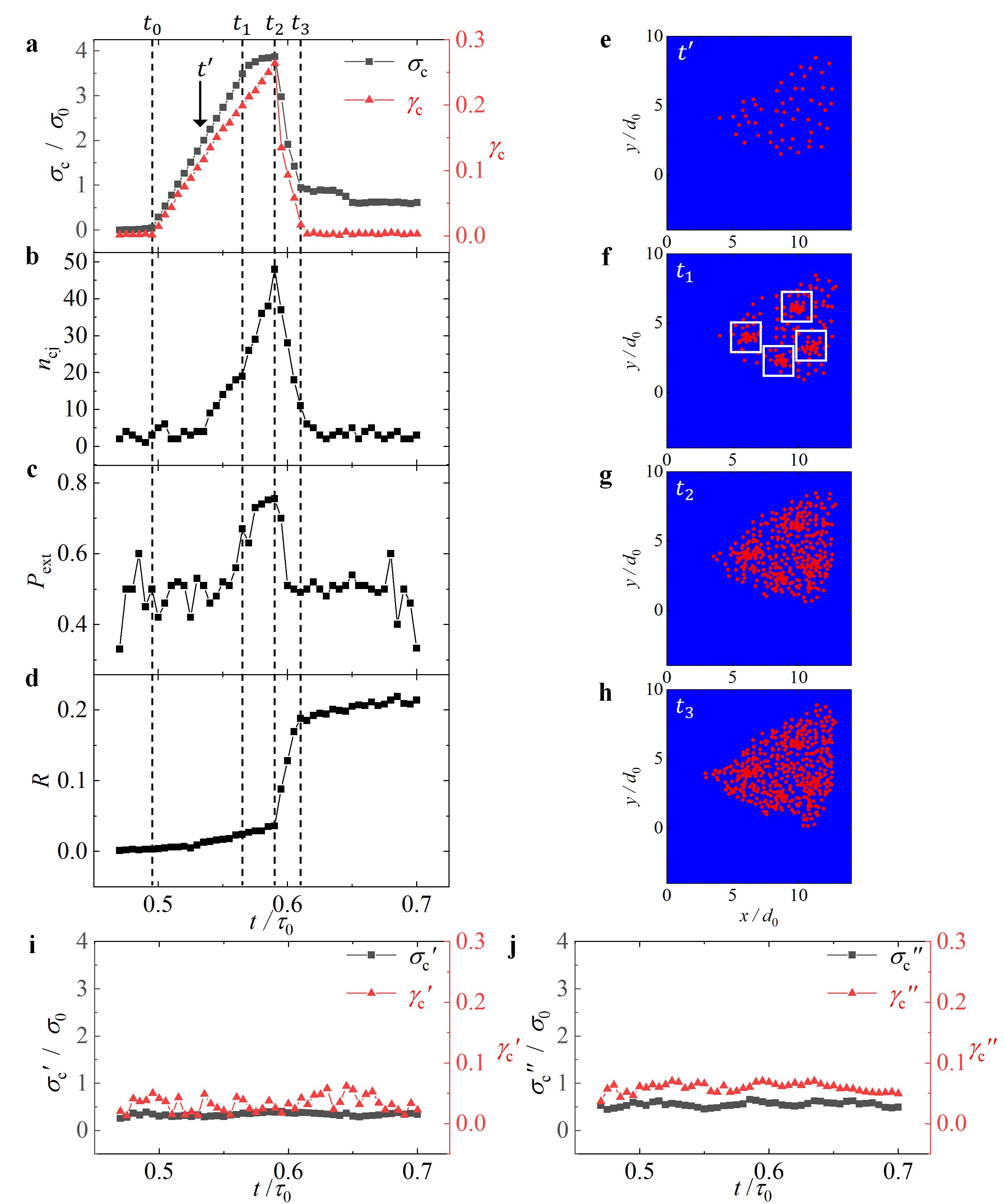}
    \caption{\label{fig:2} 
    \textbf{\textbar Results of convective cluster analysis.} 
    Evolutions of \textbf{a} $\sigma_{\mathrm{c}}$ and $\gamma_{\mathrm{c}}$, \textbf{b} $n_{\mathrm{cj}}$, \textbf{c} $P_{\mathrm{ext}}$, 
    and \textbf{d} $R$ of the convective cluster found from the dynamic region marked in Fig.~\ref{fig:1}\textbf{b} are shown. 
    Characteristic moments are noted by vertical dashed lines. 
    \textbf{e}-\textbf{h} display the spatiotemporal accumulations of cage jumps (red dots) inside the aforementioned cluster in the shear plane at $t^{\prime}$, $t_{1}$, $t_{2}$ and $t_{3}$, respectively. 
    (\textbf{i}) Evolutions of shear stress $\sigma_{\mathrm{c}}^{\prime}$ and strain $\gamma_{\mathrm{c}}^{\prime}$ of the region marked by the dashed green circle in Fig.~\ref{fig:1}\textbf{b}. 
    (\textbf{j}) Evolutions of shear stress $\sigma_{\mathrm{c}}^{\prime\prime}$ and strain $\gamma_{\mathrm{c}}^{\prime\prime}$ of the region marked by the dashed blue circle in Fig.~\ref{fig:1}\textbf{b}.}
\end{figure}

To study the formation of a dynamic region composed of cage jumps, we affinely trace back the deformation history of its boundary in flow, as illustrated in Fig.~\ref{fig:1}c. 
From a time-forward view, the boundary undergoes a convective transport in flow. 
We identify the particles located inside the time-dependent boundary every $\Delta t$ during the traceback, where $\Delta t$ is a small time interval of the order of $0.01t_{\chi}$. 
Then, we calculate the shear stress $\sigma_{\mathrm{c}}$ of the cluster encompassed by the boundary by summing the atomic shear stresses of the inside particles. 
To be specific, for particle $i$ inside the cluster, its atomic shear stress at time $t$ is given by \cite{egami3}:
\begin{equation}
	\sigma_{i,xy}(t)=\frac{1}{\Omega_{\alpha_{i}}}\sum_{j}\frac{\mathrm{d}V(r)}{\mathrm{d}r} \bigg|_{r_{ij}(t)}\frac{r_{ij,x}(t)r_{ij,y}(t)}{|\bm{r}_{ij}(t)|} \label{eq2}
	%%\alpha_{i}\in\{\mathrm{b},s\} \label{eq2}
\end{equation}
where $\bm{r}_{ij}(t)$ is the displacement between the reference particle $i$ and another particle $j$, $r_{ij,x/y}$ denote the $x/y$ component of $\bm{r}_{ij}$, 
$\Omega_{\alpha_{i}}$ represents the volume occupied by particle $i$, and the subscript $\alpha_{i}$ denotes the type of particle $i$ ($\alpha_i\in\{\mathrm{b},\mathrm{s}\}$, representing big or small). 
$\Omega_{\alpha_{i}}$ can be obtained by calculating the volume of the Voronoi cell surrounding particle $i$. 
Here for simplicity, we estimate $\Omega_{\alpha_{i}}$ by:
\begin{equation}
	\frac{\Omega_{\mathrm{b}}}{\Omega_{\mathrm{s}}}=\left(\frac{d_{\mathrm{b}}}{d_{\mathrm{s}}}\right)^{D}; N_{\mathrm{b}}\Omega_{\mathrm{b}}+N_{\mathrm{s}}\Omega_{\mathrm{s}}=V \label{eq3}
\end{equation}
where $D$ is the dimension of the system. 
Such estimation does not affect the main results and conclusions \cite{egami3}. 
Knowing $\sigma_{i,xy}(t)$, the shear stress $\sigma_{\mathrm{c}}$ of the convective cluster is found by:
\begin{equation}
	\sigma_{\mathrm{c}}(t)=\sum_{i}\sigma_{i\in F,xy}(t)\Omega_{\alpha_{i}}/V_{F} \label{eq4}
\end{equation}
where $F$ denotes the reference convective cluster, and $V_{F}$ refers to the volume of cluster $F$. 
Figure~\ref{fig:2}a gives the evolution of $\sigma_{\mathrm{c}}$ of a convective cluster, which finally gives rise to the dynamic region marked in Fig.~\ref{fig:1}b. 
It is seen that as time $t$ evolves, $\sigma_{\mathrm{c}}$ increases from nearly zero. 
We denote the starting point of the increase of $\sigma_{\mathrm{c}}$ as $t_{0}$. 
At $t>t_{0}$, $\sigma_{\mathrm{c}}$ first undergoes a linear growth until $t_{1}$, then a milder growth until $t_{2}$, and finally a sharp drop until $t_{3}$. 
Besides the shear stress, we also calculate the shear strain $\gamma_{\mathrm{c}}$ for the same cluster, and show the result in Fig.~\ref{fig:2}a. 
$\gamma_{\mathrm{c}}$ is determined by minimizing the mean-square local nonaffine displacement as suggested by Falk and Langer \cite{langer1}, and is calculated by:
\begin{equation}
	\gamma_{\mathrm{c}}(t)=\sum_{k}X_{xk}(t)Y_{yk}^{-1}(t)-\delta_{xy},k=x,y,z \label{eq5}
\end{equation}
where 
\begin{eqnarray}
	X_{xk}(t)&=&\sum_{i\in F}\left[r_{i,x}\left(t+\frac{\Delta t}{2}\right)-r_{0,x}\left(t+\frac{\Delta t}{2}\right)\right] \nonumber \\
&\times& \left[r_{i,k}\left(t-\frac{\Delta t}{2}\right)-r_{0,k}\left(t-\frac{\Delta t}{2}\right)\right] \label{eq6}
\end{eqnarray}
\begin{eqnarray}
	Y_{yk}(t)&=&\sum_{i\in F}\left[r_{i,y}\left(t-\frac{\Delta t}{2}\right)-r_{0,y}\left(t-\frac{\Delta t}{2}\right)\right] \nonumber \\
&\times& \left[r_{i,k}\left(t-\frac{\Delta t}{2}\right)-r_{0,k}\left(t-\frac{\Delta t}{2}\right)\right] \label{eq7}
\end{eqnarray}
where $\bm{r}_{0}$ denotes the position of the centroid of cluster $F$. 
Seen from Fig.~\ref{fig:2}a, $\gamma_{\mathrm{c}}$ linearly grows with $t$ at first, and again drops at $t=t_{2}$. 
The behaviors of $\sigma_{\mathrm{c}}(t)$ and $\gamma_{\mathrm{c}}(t)$ suggest that the cluster sequentially undergoes elastic deformation during $[t_{0},t_{1}]$, 
plastic deformation during $[t_{1},t_{2}]$, and yielding at $t=t_{2}$, similar to the response of amorphous solids \cite{oswald1, pete1, egelhaaf1,wangwh}. 
As we will see, the yielding process of the convective cluster is accompanied by extensive cage jumps within the cluster, which corresponds to the dynamic region marked in Fig.~\ref{fig:1}b. 
This kind of solidlike response is also found for other clusters studied in this work.

To learn more about this local response, we calculate three other quantities for the same cluster. 
Figure~\ref{fig:2} b shows the temporal density of cage jumps $n_{\mathrm{cj}}(t)$, which records the count of cage jumps in the cluster during $[t-\Delta t/2,t+\Delta t/2]$. 
$n_{\mathrm{cj}}(t)$ exhibits a smooth bump at $t_{1}$ and a peak at $t_{2}$, suggesting that the change of mechanical response is accompanied by the surge of cage jumps. 
Previous studies \cite{wang1, pete1} show that irreversible displacements of particles in sheared glassy systems are mostly along the extensional direction of shear geometry. 
Thus, we check the direction of the displacement vector $\tilde{\bm{r}}_{\mathrm{cj}}$ of every cage jump, 
and calculate the percentage $P_{\mathrm{ext}}(t)$ that the projection of $\tilde{\bm{r}}_{\mathrm{cj}}$ in the shear plane aligns with the $1^{\mathrm{st}}$ or the $3^{\mathrm{rd}}$ quadrant. 
As seen in Fig.~\ref{fig:2} c, $P_{\mathrm{ext}}(t)$ is statistically larger than $50\%$ during $[t_{0},t_{3}]$ and reaches local maximum at $t_{1}$ and $t_{2}$, 
which is consistent with previous results \cite{wang1, pete1}. 
Note that, though the evolution of the boundary is deterministic, the particles inside and outside the boundary can exchange. 
In Fig.~\ref{fig:2} d, we characterize this exchange with $R(t)=1-N_{\mathrm{res}}(t)/N_{\mathrm{res}}(t_{0})$, 
where $N_{\mathrm{res}}(t)$ is the number of particles inside the boundary at $t_{0}$ that are still inside at $t$. 
The weak exchange at $t<t_{2}$ suggests that the convective cluster can be viewed as an entity that evolves in a nearly closed manner. 
This manner is terminated by abundant rearrangements manifested by the sharp increase of $R(t)$ from $t_{2}$. 

Figure~\ref{fig:2} e-h display the spatiotemporal accumulation of cage jumps in the cluster studied in Fig.~\ref{fig:2} a-d. 
At $t<t_{1}$, cage jumps are sparse (Fig.~\ref{fig:2} e). 
At $t\approx t_{1}$, a few cage jumps have concentrated on some spots with diameter of $\sim3d_{\mathrm{0}}$, as marked in Fig.~\ref{fig:2} f. 
These spots could be the manifestation of shear transformation zones (STZ) \cite{langer1, argon1, weitz1, egami4}, as we will address later. 
During $[t_{1},t_{2}]$ (Fig.~\ref{fig:2} g), lots of cage jumps appear throughout the cluster, stemming from the spots marked in Fig.~\ref{fig:2} f. 
At $t\approx t_{3}$, the whole cluster has been occupied by cage jumps. 
The outward propagation of cage jumps from early-formed cage-jump spots shown here is consistent with the dynamic facilitation (DF) mechanism \cite{biroli2, chandler1, chandler2, berthier1}. 
Summarizing the results shown in Fig.~\ref{fig:2} a-h, 
we suggest that the emergence of the dynamic region marked in Fig.~\ref{fig:1} b is a reflection of the yielding and rearrangement of the localized elasticity embodied as a convective cluster. 

The accumulation of stress and strain in localized regions in steady shear flow and its relation to the appearance of dynamic regions are highly nontrivial. 
To further clarify the importance of these observations, we perform following contrast calculation. 
As shown in Fig.~\ref{fig:1} b, we circle two spherical regions, whose boundaries are denoted by dashed green line and blue line, respectively. 
The former is outside of any dynamic region. 
The latter contains the parts from different dynamic regions. We trace the evolutions of these two regions according to the method illustrated in Fig.~\ref{fig:1} c. 
Figure~\ref{fig:2} i and j give the results of shear stress and strain for these two regions, respectively. 
For both cases, no stress or strain accumulation is found. 
Their stresses and strains evolve steadily with fluctuations, similar to the stereotyped image of viscous liquids. 
The sharp difference between Fig.~\ref{fig:2} a and Fig.~\ref{fig:2} i, j highlights the nontriviality of the localized elastic coherency and its connection to the clustering of jump events in flow.

\subsection{Convective cluster and shear thinning}

\begin{figure}
    \includegraphics[width=\linewidth]{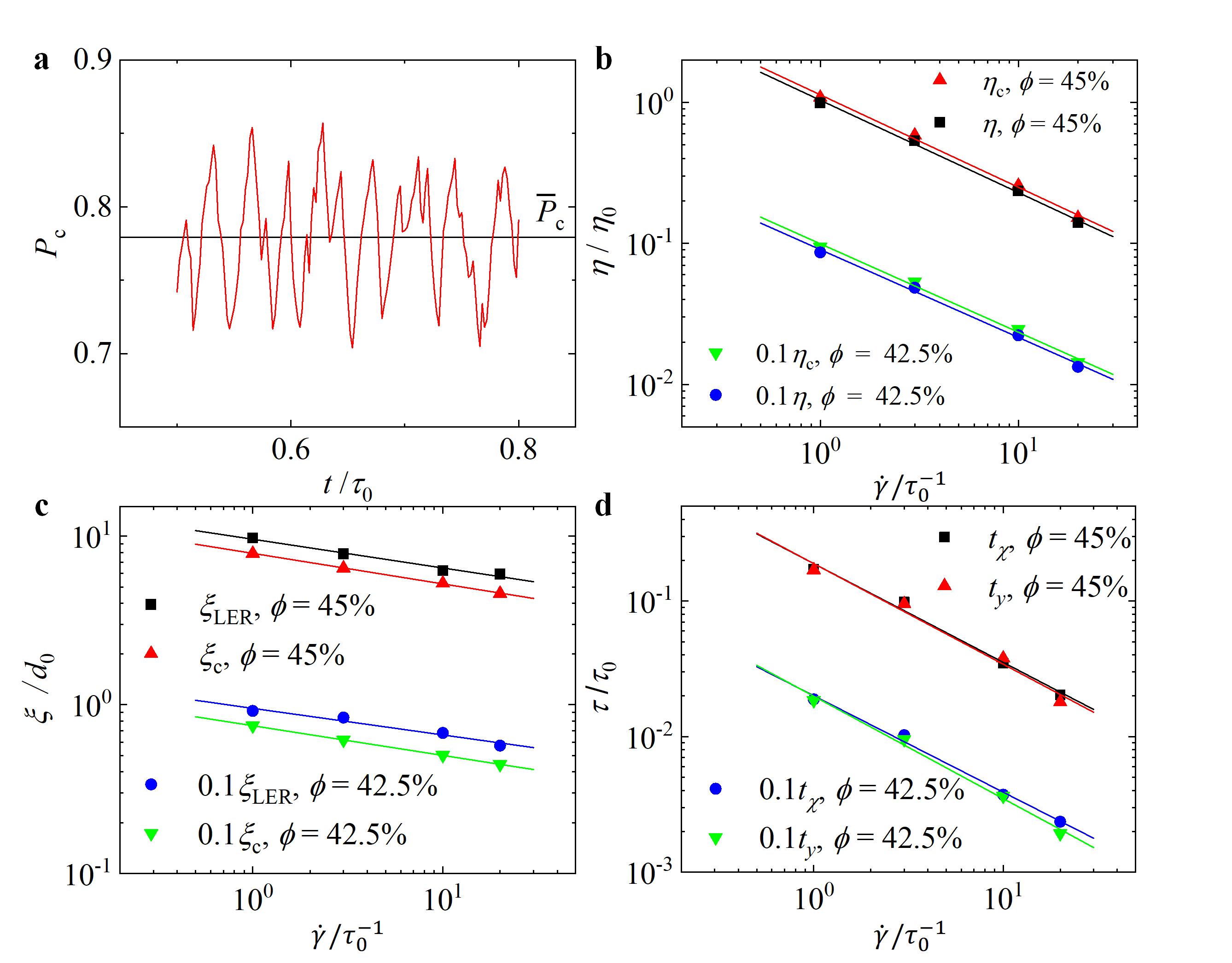}
    \caption{\label{fig:3} 
    \textbf{\textbar Convective cluster and shear thinning.} 
    (\textbf{a}) Percentage $P_{\mathrm{c}}(t)$ of particles inside convective clusters for the $\phi=45\%$ sample at $\dot{\gamma}\tau_{0}=3$. 
    The horizontal line gives its average value $\bar{P}_{\mathrm{c}}$. 
    (\textbf{b}) Comparison between the cluster-sustained viscosity $\eta_{\mathrm{c}}$ and $\eta$ calculated by eq.~\ref{eq1}. 
    Lines denote fits with $\dot{\gamma}^{-\lambda}$. 
    (\textbf{c}) Comparison between the cluster radius $\xi_{\mathrm{c}}$ and the LER radius $\xi_{\mathrm{LER}}$. 
    Lines denote fits with $\xi_{\mathrm{c}}\sim\dot{\gamma}^{-\nu_{\mathrm{c}}}$ and $\xi_{\mathrm{LER}}\sim\dot{\gamma}^{-\nu^{\prime}}$. 
    It is seen that $\nu_{\mathrm{c}}\approx\nu^{\prime}$. 
    (\textbf{d}) Comparison between the yielding time $t_{\mathrm{y}}$ and $t_{\chi}$. 
    Lines denote the power-law fits.}
\end{figure}

The preceding subsection elucidates how the yielding of convective clusters leads to the emergence of DH manifested by the clustering of cage jumps. 
Next, we will explore the relation between shear thinning and convective clusters. 
We calculate the real-time percentage $P_{\mathrm{c}}(t)$ of particles that are included in convective clusters across the whole system. 
Figure~\ref{fig:3} a shows $P_{\mathrm{c}}(t)$ for the $\phi=45\%$ BD sample at $\dot{\gamma}\tau_{0}=3$. 
It is seen that $P_{\mathrm{c}}(t)$ mildly fluctuates around its average $\bar{P}_{\mathrm{c}}$, which is $78\%$ in this case. 
In fact, $\bar{P}_{\mathrm{c}}$ is found to be about $80\%$ for all simulated flows in the shear-thinning regime. 
These results suggest that most particles are participating in the deformation-yielding process of convective clusters at a given time. 
Thus, we expect that the viscosity of the system is mainly contributed by the resistance formed by the elastic energy barrier of convective clusters. 
To test this idea, we estimate the shear viscosity of the system by the elastic stress sustained by convective clusters as:
\begin{equation}
	\eta_{\mathrm{c}}=\frac{1}{\dot{\gamma}}\bar{P}_{\mathrm{c}}\frac{\sigma_{\mathrm{c,y}}}{2} \label{eq8}
\end{equation}
where $\sigma_{\mathrm{c,y}}$ is the average yielding stress of the cluster, so that $\sigma_{\mathrm{c,y}}/2$ is the average stress sustained by the cluster during its life cycle. 
In Fig.~\ref{fig:3} b, we compare $\eta_{\mathrm{c}}$ with $\eta$ calculated by eq.~\ref{eq1} in the shear-thinning regime. 
A nice match is seen, supporting that the elasticity of convective clusters is indeed dominating in contributing shear thinning.

In Fig.~\ref{fig:3} c, we show the characteristic radius of convective clusters $\xi_{\mathrm{c}}$ as a function of $\dot{\gamma}$ in the shear-thinning regime for two BD samples. 
$\xi_{\mathrm{c}}$ is found by $\xi_{\mathrm{c}}=(3V_{\mathrm{c}}/4\pi)^{1/3}$, where $V_{\mathrm{c}}$ is the average cluster volume. 
It is seen that $\xi_{\mathrm{c}}$ shrinks with $\dot{\gamma}$ by a power-law $\xi_{\mathrm{c}}\sim\dot{\gamma}^{-\nu_{\mathrm{c}}}$. 
This is another important phenomenon for understanding the relation between localized elasticity and shear thinning. 
Inspired by the ``solids-that-flow'' picture of supercooled liquids \cite{dyre1, dyre2, dyre3}, 
we give the following picture for local structural rearrangement in the shear-thinning regime \cite{wang1}: 
Before the rearrangement takes place, most particle motion is vibrational, and jump events are relatively rare. 
Such solidlike behavior cannot be global for liquids. 
It only happens within certain length, denoted as $l_{\mathrm{sol}}$. 
In our case, $l_{\mathrm{sol}}$ is just the length scale of convective cluster $\xi_{\mathrm{c}}$. 
Set $l_{0}$ the characteristic length of a jump event. 
Within the solidlike region, the number of possible locations for jump events is about $N_{\mathrm{j}}{\sim}(\xi_{\mathrm{c}}/l_{0})^{D}$. 
Considering that the local configuration will thoroughly rearrange after the yielding and avalanche of a convective cluster, 
we can estimate the relaxation time of local structure $\tau_{\mathrm{LS}}$ by the life cycle of the convective cluster $\tau_{\mathrm{LS}}\sim\gamma_{\mathrm{c,y}}/\dot{\gamma}$, 
where $\gamma_{\mathrm{c,y}}$ is the average yielding strain of cluster. 
Then, the average time between two time-adjacent jumps is given by $\tau_{\mathrm{LS}}/N_{\mathrm{j}}$. 
To keep the solidity within the cluster, $\tau_{\mathrm{LS}}/N_{\mathrm{j}}$ should be equal to $l_{\mathrm{sol}}/c$, where $c$ is the transverse sound speed. 
Therefore, we have $\xi_{\mathrm{c}}^{D+1}\approx cl_{0}^{D}\gamma_{\mathrm{c,y}}/\dot{\gamma}\propto\gamma_{\mathrm{c,y}}/\dot{\gamma}$. 
To connect $\xi_{\mathrm{c}}$ and shear thinning, we take notice of the relation $\gamma_{\mathrm{c,y}}\sim\dot{\gamma}^{1-\lambda}$, 
which can be derived from $\eta_{\mathrm{c}}\sim\sigma_{\mathrm{c,y}}/\dot{\gamma}\sim\dot{\gamma}^{-\lambda}$ and $\sigma_{\mathrm{c,y}}\propto\gamma_{\mathrm{c,y}}$. 
Consequently, we have $\xi_{c}^{D+1}\sim\dot{\gamma}^{-\lambda}$. 
Recalling the observation of $\xi_{\mathrm{c}}\sim\dot{\gamma}^{-\nu_{\mathrm{c}}}$, we finally find the following scaling law \cite{wang1}:
\begin{equation}
	\nu_{\mathrm{c}}=\frac{\lambda}{D+1} \label{eq9}
\end{equation}
We compare the values of $\nu_{\mathrm{c}}$ obtained by fitting the data shown in Fig.~\ref{fig:3} c and the values of $\frac{\lambda}{D+1}$ in Table~\ref{table:1}. 
It is seen that the relative difference between these two exponents is only about $5\%$. 
The nice validity of eq.~\ref{eq9} and the observation that $\xi_{\mathrm{c}}$ shrinks with $\dot{\gamma}$ further elucidate the essential role of localized elasticity in shear thinning. 
Apparently, shear thinning is related to the fact that $\gamma_{\mathrm{c,y}}$ does not increase linearly with $\dot{\gamma}$, 
but with a weaker relation $\gamma_{\mathrm{c,y}}\sim\dot{\gamma}^{1-\lambda}$. 
According to the aforementioned picture, the weaker relation  $\gamma_{\mathrm{c,y}}\sim\dot{\gamma}^{1-\lambda}$ must be accompanied with the shrinkage of $\xi_{\mathrm{c}}$. 
Note that, if $\xi_{\mathrm{c}}$ does not shrink with $\dot{\gamma}$, the jump events inside a convective cluster will be too frequent. 
In this case, transverse phonons will not have enough time to traverse the whole cluster and, consequently, the elasticity cannot be sustained within the whole cluster. 

According to the procedure of finding convective clusters from dynamic regions illustrated in Fig.~\ref{fig:1} c, 
$\xi_{\mathrm{c}}$ is also the characteristic length of dynamic regions. 
Therefore, eq.~\ref{eq9} bridges shear thinning and DH through localized elasticity. 
To further confirm this picture, we compare $t_{\chi}$ with the average yielding time $t_\mathrm{y}=\langle t_{2}-t_{0}\rangle$ of convective clusters in the shear-thinning regime. 
As shown in Fig.~\ref{fig:3} d, $t_{\chi}$ nicely agrees with $t_{\mathrm{y}}$. 
This result is consistent with our viewpoint that the emergence of prominent DH is intimately correlated with the massive yielding of elastic clusters.

In our previous work \cite{wang1}, we propose the concept of LER within which the response to shear is elastic. 
Conceptually, the convective cluster is the spatiotemporal representation of LER. 
In that work, we evaluate the characteristic radius of LER $\xi_{\mathrm{LER}}$ by analyzing the distortion of $g(\bm{r})$ \cite{wang1}. 
The results of $\xi_{\mathrm{LER}}$ are also plotted in Fig.~\ref{fig:3} c. 
It is seen that $\xi_{\mathrm{LER}}$ and $\xi_{\mathrm{c}}$ behave highly consistently, with $\xi_{\mathrm{c}}\approx0.8\xi_{\mathrm{LER}}$ for all conditions. 
The agreement between $\xi_{\mathrm{LER}}$ and $\xi_{\mathrm{c}}$ is noticeable, 
especially considering that $\xi_{\mathrm{LER}}$ is found from the 2-point correlation \cite{wang1}, 
which is commonly considered to be not sensitive to DH features \cite{glotzer1}. 

\begin{table}
	\caption{\textbf{\textbar $\frac{\lambda}{D+1}$ and $\nu_{\mathrm{c}}$ for different model systems.}}\label{table:1}%
	\begin{ruledtabular}
    \begin{tabular}{ccc}
		% \toprule
		Model System & $\frac{\lambda}{D+1}$  & $\nu_{\mathrm{c}}$ \\
		\hline
		3D BD, $\phi=45\%$    & 0.166   & 0.174    \\
		3D BD, $\phi=42.5\%$    & 0.158   & 0.163    \\
		3D MD    & 0.182   & 0.176    \\
		2D MD    & 0.253   & 0.247    \\
		% \botrule
	\end{tabular}
    \end{ruledtabular}
\end{table}

\subsection{Yielding mechanism of convective cluster}\label{subsec2.3}

\begin{figure}
    \includegraphics[width=\linewidth]{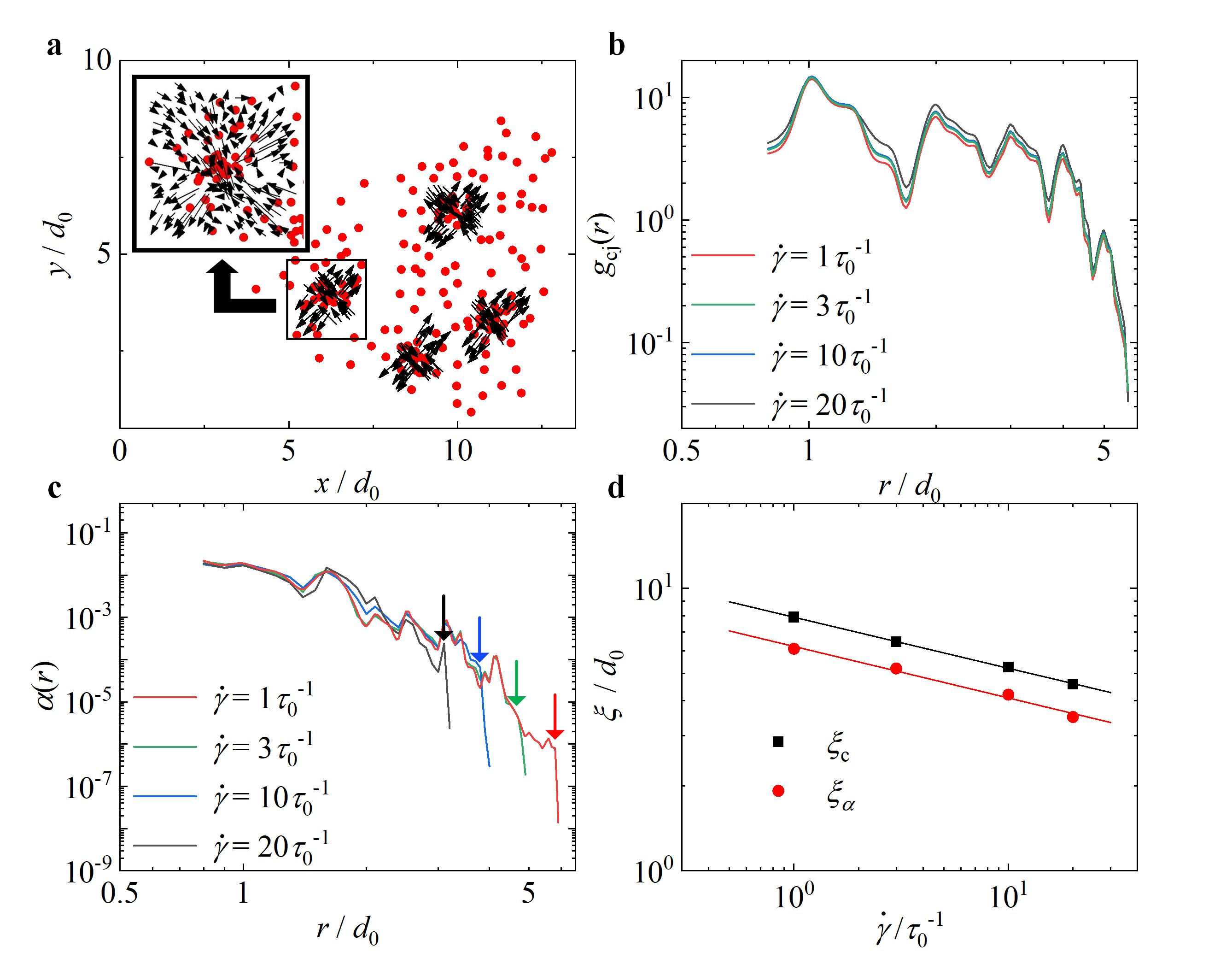}
    \caption{\label{fig:4} 
    \textbf{\textbar Yielding mechanism of convective cluster.} 
    (\textbf{a}) Projection of PHM in the shear plane for the convective cluster studied in Fig.~\ref{fig:2} \textbf{a}-\textbf{h}. 
    Cage jumps happening during $[t_{0},t_{1}]$ are also shown (red dots). 
    Here, only the projections with significant magnitude are retained for clarity. 
    The inset is the enlarged view of the spot marked by the rectangle and shows all PHM projections around the marked cage-jump spot. 
    The excellent coincidence between PHM-predicted STZs and cage-jump spots is seen. 
    \textbf{b} and \textbf{c} respectively show the orientation-averaged $g_{\mathrm{cj}}(r)$ and the anisotropic factor $\alpha(r)$ of the $\phi=45\%$ BD sample at different $\dot{\gamma}$. 
    (\textbf{d}) Comparison between the radius of $\alpha(r)$ $\xi_{\alpha}$ and the cluster radius $\xi_{\mathrm{c}}$ for the $\phi=45\%$ BD sample. Lines denote the power-law fits.}
\end{figure}

The above discussion highlights the necessity of understanding the microscopic mechanism of the yielding of convective clusters in understanding shear thinning. 
Considering the similarity between the responses of convective clusters and amorphous solids to shear, one may expect that the yielding of convective clusters is initiated by STZs. 
STZs are known to relate to the undeformed configuration \cite{manning1}. 
To search for STZs, we first isolate the cluster studied in Fig.~\ref{fig:2} a-h by fixing the particles outside the cluster, 
and then calculate the pseudoharmonic modes (PHM) \cite{lerner1} with reference to the inherent structure of the cluster at $t_{0}$. 
The PHM, defined based on the harmonic approximation of a glass's energy, is proven to be an effective indicator for STZ \cite{lerner1}. 
In Fig.~\ref{fig:4} a, we show the modes that indicate potential STZs in the cluster. 
The aggregation of these eigenvectors is featured by the geometry consistent with the shear: extensional in the $1^{\mathrm{st}}$ and the $3^{\mathrm{rd}}$ quadrants, and compressional in the other two. 
The cage jumps happening during $[t_{0},t_{1}]$ are also plotted, as we do in Fig.~\ref{fig:2} f. 
It is seen that the cage-jump spots marked in Fig.~\ref{fig:2} f coincide the loci of STZs predicted by PHM analysis. 
The existence of STZ inside convective clusters suggests that the dynamics of sheared supercooled liquids manifests an interesting spatial hierarchy of 
\emph{liquidlike background} \textemdash \emph{solidlike medium (LER)} \textemdash \emph{liquidlike spots (STZ)} from large length scales to small length scales.

Local plastic events trigger anisotropic stress correlation in surrounding elastic medium \cite{picard1}. 
This mechanism, which has been incorporated into elastoplastic models for the rheology of amorphous solids \cite{barrat1}, 
was introduced to study the DH in equilibrium supercooled liquids by means of DF recently \cite{liu1, biroli1}. 
The development of cage jumps shown in Fig.~\ref{fig:2} e-h exhibits DF-like feature. 
To clarify the relation between the localized elasticity and the likely DF, for a specific convective cluster, we calculate the pair correlation of cage jumps as \cite{liu1}:
\begin{equation}
	g_{\mathrm{cj}}(\boldsymbol{r})=\frac{N_{\mathrm{cj}}^{\prime}}{N}\langle\sum_{j,k}\delta\left(\boldsymbol{r}-\left[\boldsymbol{r}_{\mathrm{s},k}(t_1:t_2)-\boldsymbol{r}_{\mathrm{c},j}(t_1)\right]\right)\rangle \label{eq10}
\end{equation}
where $\bm{r}_{\mathrm{c},j}(t_{1})$ denotes the position of the $j^{\mathrm{th}}$ cage jump appearing within $[t_{1}-\Delta t/2,t_{1}+\Delta t/2]$ inside the cluster, 
$\bm{r}_{\mathrm{s},k}(t_{1}\colon t_{2})$ denotes the position of the $k^{\mathrm{th}}$ cage jumps appearing within $[t_{1},t_{2}]$ in the whole system, 
$N$ is the number of particles, and $N_{\mathrm{cj}}^{\prime}$ is the count of cage jumps appearing within $[t_{0},t_{1}]$ in the whole system. 
$g_{\mathrm{cj}}(\boldsymbol{r})$ evaluates the facilitation effect during $[t_{1},t_{2}]$, within which abundant cage jumps emerge. 
Figure~\ref{fig:4} b gives the orientation-averaged $g_{\mathrm{cj}}(r)$ for the $\phi=45\%$ BD sample. 
It is seen that $g_{\mathrm{cj}}(r)$ is not sensitive to $\dot{\gamma}$. 
Considering the anisotropy of the strain field of plastic events, it is the anisotropy of $g_{\mathrm{cj}}(\boldsymbol{r})$, 
rather than the orientation-averaged radial feature, that better measures the elastic component of $g_{\mathrm{cj}}(\boldsymbol{r})$ \cite{liu1}. 
Thus, we calculate the anisotropic factor of $g_{\mathrm{cj}}(\boldsymbol{r})$ by $\alpha(r)=\langle\left[g_{\mathrm{cj}}(\boldsymbol{r})/g_{\mathrm{cj}}(r)-1\right]^2\rangle_\Omega $, 
where $\langle...\rangle_{\Omega}$ represents average over solid angle. 
As seen in Fig.~\ref{fig:4} c, the amplitude of $\alpha(r)$ does not depend on $\dot{\gamma}$, whereas its spatial range clearly shrinks with $\dot{\gamma}$. 
We identify the truncation radius of $\alpha(r)$, denoted as $\xi_{\alpha}$, by the point at which $\alpha(r)$ undergoes a sudden change to a much steeper profile, as marked by arrows in Fig.~\ref{fig:4} c. 
Figure ~\ref{fig:4} d displays the result of $\xi_{\alpha}$. $\xi_{\mathrm{c}}$ is also plotted for comparison. 
$\xi_{\alpha}$ and $\xi_{\mathrm{c}}$ are seen to be be highly consistent with each other manifested by $\xi_{\alpha}\approx0.8\xi_{\mathrm{c}}$. 
This agreement confirms the localized nature of the transient elasticity, 
and reveals the importance of elasticity-mediated interaction in facilitating the avalanche of cage jumps and the yielding of convective clusters.

\subsection{Atomic system and 2D system}\label{subsec2.4}

\begin{figure}
    \includegraphics[width=\linewidth]{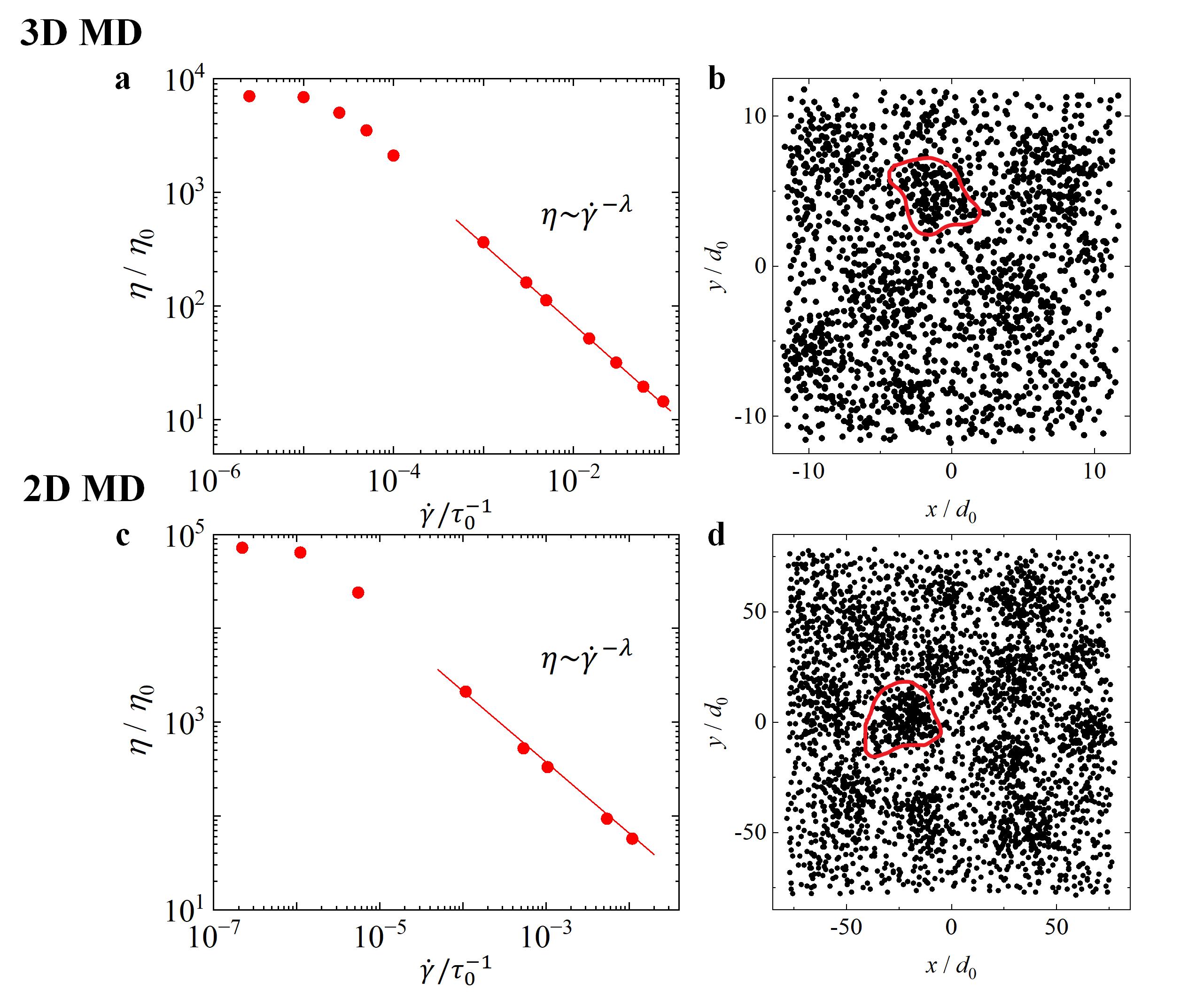}
    \caption{\label{fig:5} 
    \textbf{\textbar Shear thinning and dynamic heterogeneity of MD systems.} 
    (\textbf{a}) Shear viscosity of the 3D MD system. 
    (\textbf{b}) Projection of cage-jump events in the shear plane accumulated within a time interval $t_{\chi}$ for the 3D MD system at $\dot{\gamma}\tau_0=0.015$. 
    (\textbf{c}) Shear viscosity of the 2D MD system. 
    (\textbf{d}) Cage-jump events accumulated within a time interval $t_{\chi}$ for the 2D MD system at $\dot{\gamma}\tau_0=0.001$. 
    Solid lines in \textbf{a} and \textbf{c} denote fits with $\dot{\gamma}^{-\lambda}$ in the shear-thinning regime. 
    Red solid curves in \textbf{b} and \textbf{d} denote the boundaries of selected cage-jump clusters.}
\end{figure}

In the above subsections, we establish the relation between shear thinning and DH based on the 3D BD systems, which represent colloidal suspensions. 
One may be interested in if our conclusion also holds for atomic systems. 
Additionally, the scaling law given by eq.~\ref{eq9} calls for the validation with other dimensions. 
For these purposes, we perform MD simulations in 3D and 2D, as introduced in Methods. 
For both MD samples, shear thinning is remarkable and can be described with a power law $\eta{\sim}\dot{\gamma}^{-\lambda}$ at high enough shear rates, as shown in Fig.~\ref{fig:5} a and c. 
At the microscopic level, the DH phenomenon, which is manifested by the spatiotemporal clustering of cage jumps, is clearly seen in both MD samples, as shown in Fig.~\ref{fig:5} b and d. 
These observations are similar to those of the BD systems. 

\begin{figure}
    \includegraphics[width=\linewidth]{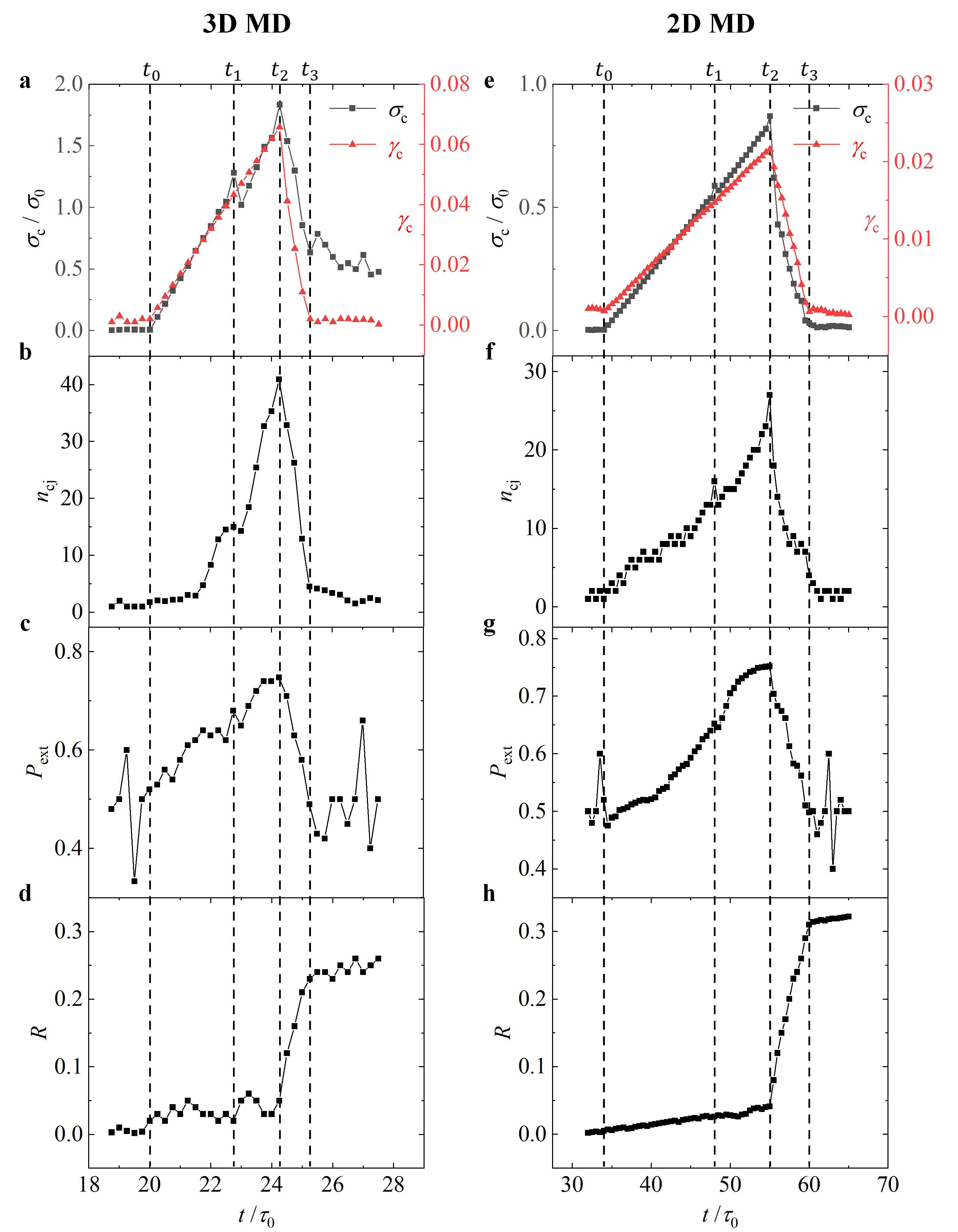}
    \caption{\label{fig:6} 
    \textbf{\textbar Convective cluster analysis for MD systems.} 
    The left column shows evolutions of \textbf{a} $\sigma_{\mathrm{c}}$ and $\gamma_{\mathrm{c}}$, \textbf{b} $n_{\mathrm{cj}}$, \textbf{c} $P_{\mathrm{ext}}$, 
    and \textbf{d} $R$ of the convective cluster found from the dynamic region marked in Fig.~\ref{fig:5}\textbf{b}. 
    The right column shows evolutions of \textbf{e} $\sigma_{\mathrm{c}}$ and $\gamma_{\mathrm{c}}$, \textbf{f} $n_{\mathrm{cj}}$, \textbf{g} $P_{\mathrm{ext}}$, 
    and \textbf{h} $R$ of the convective cluster found from the dynamic region marked in Fig.~\ref{fig:5}\textbf{d}. 
    Characteristic moments are noted by vertical dashed lines.}
\end{figure}

To examine the relation between shear thinning and DH for MD samples, we identify the convective clusters and perform the cluster analysis for two MD samples. 
Figure~\ref{fig:6} shows typical results of the cluster analysis for both MD samples. 
Here, the left column and the right column display the results of the 3D sample and the 2D sample, respectively. 
The two clusters studied in Fig.~\ref{fig:6} are found from the dynamic regions marked in Fig.~\ref{fig:5} b and d. 
In calculating $\sigma_{\mathrm{c}}(t)$, we employ the iso-configurational ensemble (ICE) method \cite{harrowell1} to enhance the statistics with $50$ ICE replicants for each cluster. 
It is seen that the results of MD samples are qualitatively consistent with those of the BD sample shown in Fig.~\ref{fig:2} a-d, 
suggesting that the convective clusters in MD systems also undergo the three-step process: elastic deformation, plastic yielding, and sharp disintegration. 
Moreover, we check the spatiotemporal accumulations of cage jumps for these two clusters, as what we do in Fig.~\ref{fig:2} e-h. 
Meanwhile, we also perform the PHM calculations on them. 
The results confirm the emergence of cage-jump spots, which can be the manifestation of STZs, at the end of the elastic deformation stage in MD systems. 
The small overshoots of $\sigma_{\mathrm{c}}$ at $t_{1}$ and $t_{2}$ shown in Fig.~\ref{fig:6} a and e could be attributed to 
the over-distortion of the corresponding cage along the extensional direction when a particle is about to jump.

\begin{figure}
    \includegraphics[width=\linewidth]{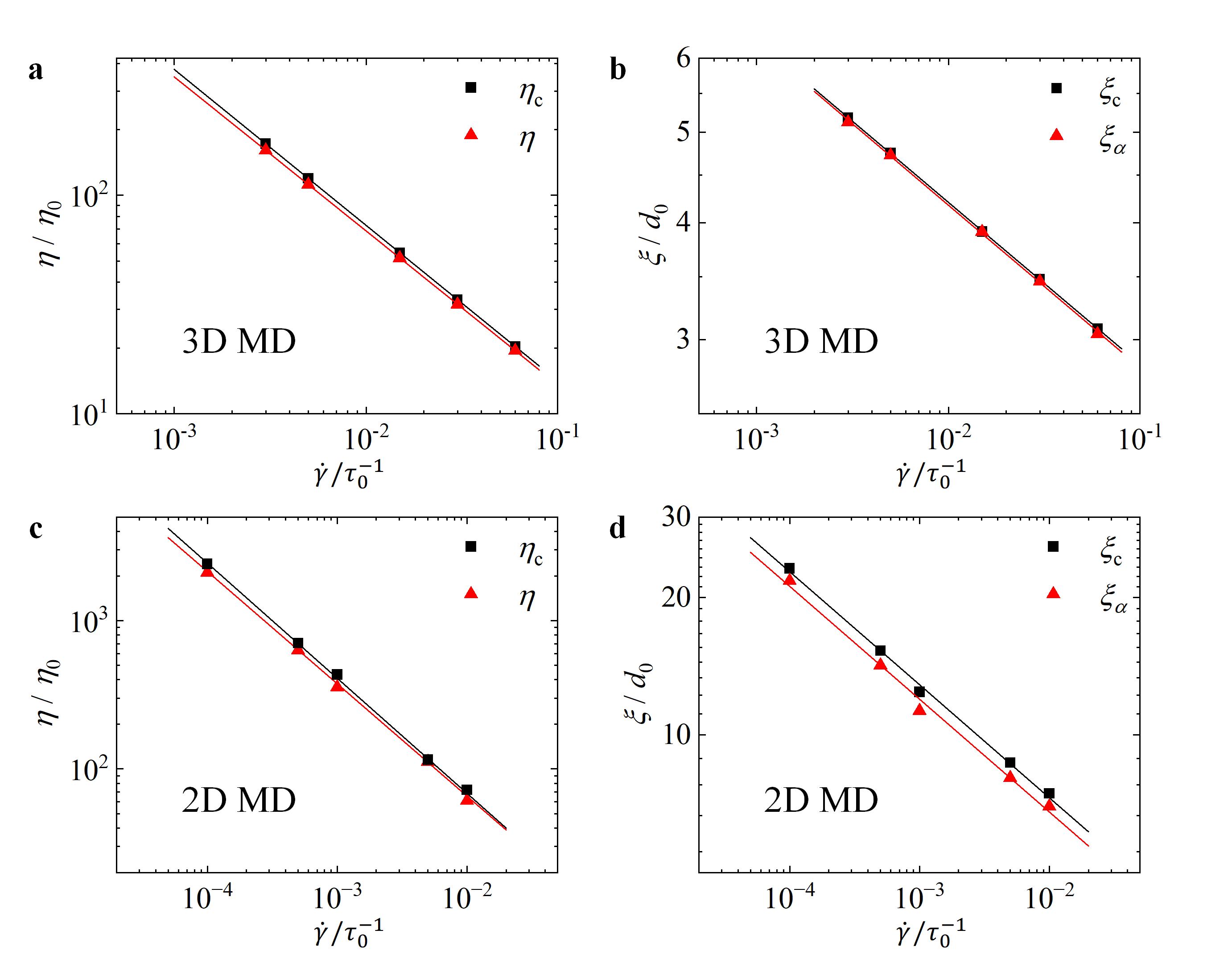}
    \caption{\label{fig:7} 
    \textbf{\textbar Convective cluster and shear thinning in MD systems.} 
    \textbf{a} and \textbf{b} compare the cluster-sustained viscosity $\eta_{\mathrm{c}}$ and $\eta$ calculated by eq.~\ref{eq1} for the 3D system and the 2D system, respectively. 
    \textbf{c} and \textbf{d} compare the cluster radius $\xi_{\mathrm{c}}$ and the radius of $\alpha(r)$ $\xi_{\alpha}$ for the 3D system and the 2D system, respectively. 
    Lines denote power-law fits in all panels.}
\end{figure}

We calculate the viscosity sustained by cluster $\eta_{\mathrm{c}}$ and the radius of cluster $\xi_{\mathrm{c}}$ in the shear-thinning regime for two MD samples. 
As seen from Fig.~\ref{fig:7} a and c, $\eta_{\mathrm{c}}$ nicely matches $\eta$ calculated by eq.~\ref{eq1}, confirming the importance of the localized elasticity in resisting external flow. 
The results of $\xi_{\mathrm{c}}$ are shown in Fig.~\ref{fig:7} b and d. 
Again, $\xi_{\mathrm{c}}$ shrinks with $\dot{\gamma}$ following a power-law form $\xi_{\mathrm{c}}\sim\dot{\gamma}^{-\nu_{\mathrm{c}}}$. 
Note that, the scaling relation given in eq.~\ref{eq9} critically depends on dimension. 
As seen in Table~\ref{table:1}, both the 2D sample and the 3D sample well follow the scaling relation, thereby convincingly demonstrating its validity. 
We also plot the results of truncation radius of $\alpha(r)$ ($\xi_{\alpha}$) in Fig.~\ref{fig:7} b and d. 
For both MD samples, $\xi_{\alpha}$ is approximately equal to $\xi_{\mathrm{c}}$, implying the crucial role of the elasticity-mediated interaction in our picture.

With Fig.~\ref{fig:5} to Fig.~\ref{fig:7}, we show that the major conclusions still hold in MD systems and in 2D.

\section{Conclusion}\label{sec3}
In this work, we explore the connection between two universal dynamic phenomena in supercooled liquids, namely, shear thinning at the macroscopic level and DH at the microscopic level. 
We demonstrate the intermediate role of localized elasticity in this connection: 
It is the major source of viscosity that shear-thins, and its yielding gives rise to dynamic regions that constitute the prominent manifestation of DH. 
The yielding of the localized elasticity is initiated by STZs and facilitated by the anisotropic elasticity-mediated interaction. 
With these observations, we quantitatively link DH and shear thinning by a simple scaling relation $\nu_{\mathrm{c}}=\frac{\lambda}{D+1}$. 
These results pave a way for the modeling of the nonlinear rheology of supercooled liquids by elucidating the behaviors of localized elasticity.

\section{Methods}\label{sec4}
\subsection{Simulation systems}\label{subsec4.1}
The details of the three simulation systems are as follows. 
(i) The BD system is the same as the one we employed in our previous study \cite{wang1}: 
a 3D binary mixture of $N_\mathrm{s}=4000$ small particles and $N_\mathrm{b}=16000$ big particles (diameter ratio $d_\mathrm{s}/d_\mathrm{b}=2/3$) interacting via the hard-sphere Yukawa potential. 
The hard core is represented by the potential-free algorithm \cite{brady1, heyes1}, 
and the Yukawa part is expressed as $V(r)=K\exp\left[-z\left(r-d_{ij}\right)\right]/\left(r/d_{ij}\right)$ for $r\geq d_{ij}\equiv(d_{i}+d_{j})/2$, 
where $d_{i}$ is the diameter of particle $i$, 
and the parameters $z=4.86/d_\mathrm{b}$ and $K=9.69k_\mathrm{B}T$ ($k_\mathrm{B}$ is the Boltzmann constant) are determined from 
small-angle neutron scattering results of a charged colloidal suspension \cite{wang2}. 
$V(r)$ is truncated at $r=5d_{\mathrm{b}}$. 
The units of length and time are, respectively, set by $d_0=d_\mathrm{b}$ and $\tau_0=d_{\mathrm{b}}^2/D_0$ where $D_0$ is the self-diffusion coefficient of the big particle at dilute limit. 
Two volume fractions of particles, $\phi=45\%$ and $42.5\%$, are adopted. 
The long-time self-diffusion coefficient at $\phi=42.5\%$ is found to be $0.06D_0$, which is well below the dynamical criterion for freezing of colloidal suspensions \cite{lowen1}, 
suggesting that the BD samples are in supercooled state. 
(ii) For the 3D MD system, we adopt the Kob-Andersen liquid \cite{kob1}: 
a 3D binary mixture of $N_{\mathrm{s}}=3300$ small particles and $N_{\mathrm{b}}=13200$ big particles interacting via 
the Lennard-Jones (LJ) potential $V(r)=4\epsilon_{\alpha\beta}\left[\left(\sigma_{\alpha\beta}/r\right)^{12}-\left(\sigma_{\alpha\beta}/r\right)^6\right]$ with $\alpha,\beta\in\{\mathrm{b},\mathrm{s}\}$. 
Big particles and small particles are with the same mass $m$. 
LJ parameters are given by $\epsilon_{\mathrm{bb}}=1$, $\sigma_{\mathrm{bb}}=1$, $\epsilon_{\mathrm{bs}}=1.5$, $\sigma_{\mathrm{bs}}=0.8$, $\epsilon_{\mathrm{ss}}=0.5$, 
and $\sigma_{\mathrm{ss}}=0.88$. 
$V(r)$ is truncated at $r=2.5\sigma_{\mathrm{bb}}$. 
The units of length and time are, respectively, set by $d_0=\sigma_{\mathrm{bb}}$ and $\tau_0=\sqrt{m\sigma_{\mathrm{bb}}^2/\epsilon_{\mathrm{bb}}}$. 
The particle number density is set to $\rho=1.2$. 
The temperature is set to $k_\mathrm{B}T/\epsilon_\mathrm{bb}=0.45$, slightly higher than the mode-coupling-theory temperature $k_\mathrm{B}T_\mathrm{MCT}/\epsilon_\mathrm{bb}=0.435$. 
(iii) In addition to the above two 3D systems, we also simulate a 2D MD system to test the scaling law. 
The 2D system is the same as the one used in a previous simulation study of shear thinning \cite{tanaka2}: 
a binary mixture of $N_\mathrm{s}=10^4$ small disks and $N_\mathrm{b}=10^4$ big particles interacting via a soft-core potential $V(r)=\epsilon(\sigma_{\alpha\beta}/r)^{12}$, 
where $\sigma_{\alpha\beta}=(\sigma_{\alpha}+\sigma_{\beta})/2$ with $\alpha,\beta\in\{\mathrm{b},\mathrm{s}\}$. 
$V(r)$ is truncated at $r=4.5\sigma_{\mathrm{s}}$. 
The mass and size ratios are set to $m_\mathrm{b}/m_\mathrm{s}=2$ and $\sigma_{\mathrm{b}}/\sigma_{\mathrm{s}}=1.4$. 
The units of length and time are, respectively, set by $d_0=\sigma_{\mathrm{s}}$ and $\tau_{0}=\sqrt{m_{\mathrm{s}}\sigma_{\mathrm{s}}^{2}/\epsilon}$. 
The particle number density is set to $\rho=0.8$. 
The temperature is set to $k_\mathrm{B}T/\epsilon=0.526$. 
Here, both the 3D MD and 2D MD systems are simulated with the LAMMPS package \cite{lammps}. 

As for the units of stress $\sigma_{0}$ and viscosity $\eta_{0}$, we adopt $\sigma_{0}=k_{\mathrm{B}}T/d_{0}^{3}$ and $\eta_0=k_{\mathrm{B}}T\tau_0/d_0^3$ in BD simulations; 
$\sigma_{0}=\epsilon_{\mathrm{bb}}/d_{0}^{3}$ and $\eta_0=\epsilon_\mathrm{bb}\tau_0/d_0^3$ in 3D MD simulations; 
$\sigma_0=\epsilon/d_0^2$ and $\eta_0=\epsilon\tau_0/d_0^2$ in 2D MD simulations. 

In the main text, we use $\Delta t$ as the time interval for calculations of many quantities. 
Its value is set as follows: For BD simulations,$\Delta t=0.005\tau_0$ for $\dot{\gamma}=1\tau_0^{-1}$ and $3\tau_0^{-1}$; $\Delta t=0.002\tau_0$ for $\dot{\gamma}=10\tau_0^{-1}$; 
$\Delta t=0.001\tau_0$ for $\dot{\gamma}=20\tau_0^{-1}$. 
For 3D MD simulations, $\Delta t=0.2\tau_0$ for $\dot{\gamma}=0.003\tau_0^{-1}$, $\Delta t=0.15\tau_0$ for $\dot{\gamma}=0.005\tau_0^{-1}$, 
$\Delta t=0.1\tau_0$ for $\dot{\gamma}=0.015\tau_0^{-1}$ and $0.03\tau_0^{-1}$, and $\Delta t=0.05\tau_0$ for $\dot{\gamma}=0.06\tau_0^{-1}$. 
For 2D MD simulations, $\Delta t=2\tau_0$ for $\dot{\gamma}=10^{-4}\tau_0^{-1}$, $\Delta t=\tau_0$ for $\dot{\gamma}=5\times10^{-4}\tau_0^{-1}$, 
$\Delta t=0.5\tau_0$ for $\dot{\gamma}=10^{-3}\tau_0^{-1}$, $\Delta t=0.1\tau_0$ for $\dot{\gamma}=5\times10^{-3}\tau_0^{-1}$ and $\Delta t=0.05\tau_0$ for $\dot{\gamma}=10^{-2}\tau_0^{-1}$.

\subsection{Identification of cage jumps and demarcation of dynamic regions}\label{subsec4.2}

\begin{figure}
    \includegraphics[width=\linewidth]{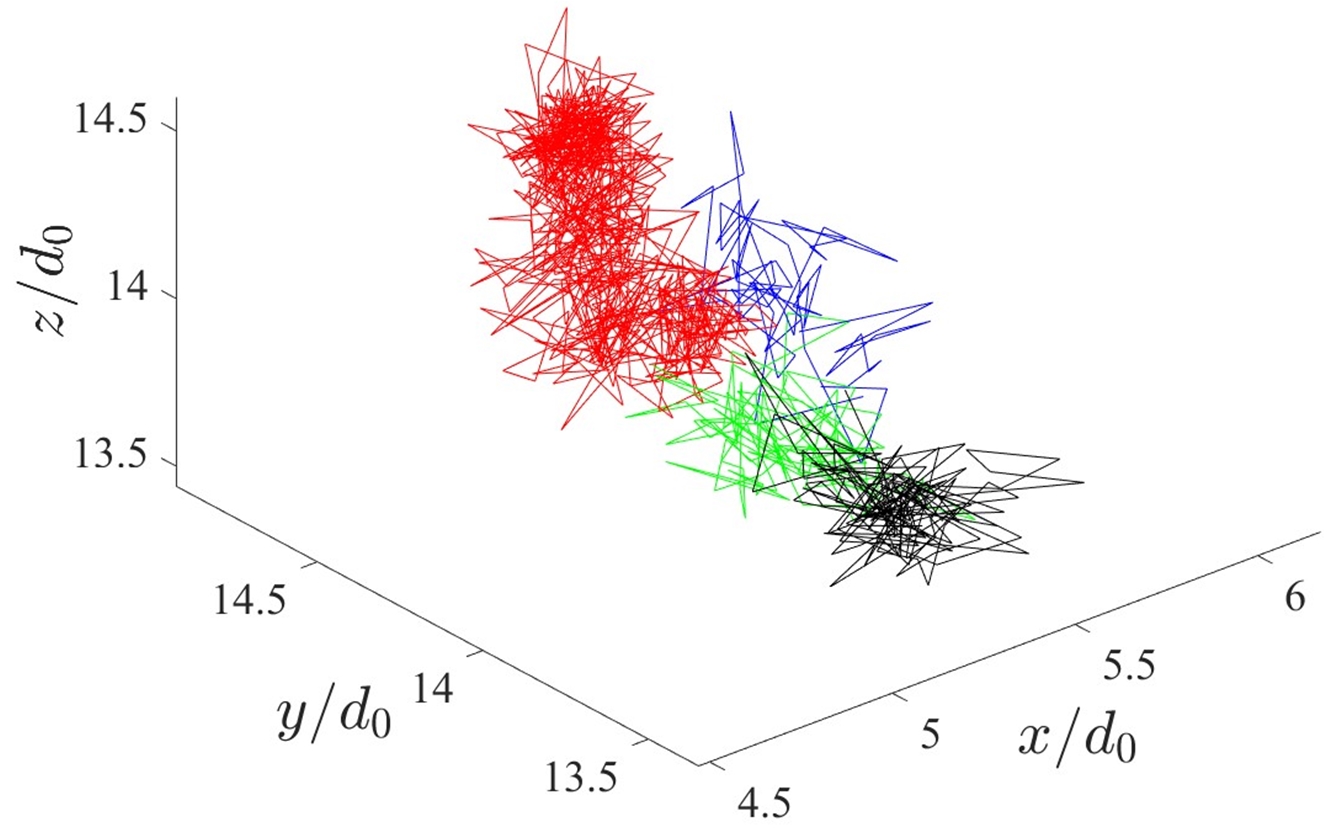}
    \caption{\label{fig:8} 
    \textbf{\textbar Trajectory of a reference particle.} The color of trajectory changes at every cage jump. Four segments are shown here.}
\end{figure}

We identify the irreversible nonaffine displacements of particles, noted as cage jumps, following the algorithm given by Candelier et al. \cite{biroli2, biroli3}. 
Considering the existence of external flow, we adopted the nonaffine particle displacement $\tilde{\boldsymbol{r}}(t)$ defined by Yamamoto and Onuki \cite{yamamoto1} as the particle trajectory. 
First, we split the trajectory of a particle $S(t)_{t\in[0,T]}$ into two sets of successive points, respectively denoted as $S_{1}$ and $S_{2}$, at an arbitrary cut time $t_{\mathrm{c}}$. 
One can evaluate how well separated are these two sets of points by
\begin{equation}
	p(t_\mathrm{c})=\zeta(t_\mathrm{c})\left[\langle d_1^2(t_2)\rangle_{t_2\in S_2}\langle d_2^2(t_1)\rangle_{t_1\in S_1}\right]^{1/2} \label{eq11}
\end{equation}
where $\zeta(t_\mathrm{c})=\sqrt{t_\mathrm{c}/T\times(1-t_\mathrm{c}/T)}$, 
$d_k(t_i)$ is the distance between the point at time $t_i$ and the center of mass of the subset $S_k$. 
The average $\langle...\rangle_{S_{k}}$ is over the subset $S_k$. 
A cage jump event is defined at $t_{\mathrm{c}}$ if $p(t_{\mathrm{c}})$ is maximal. 
Then, by iteratively repeating this procedure for every sub-trajectory until $p_{\max}(t_{\mathrm{c}})$ is smaller than a threshold $l_{\mathrm{c}}^2$ ($l_{\mathrm{c}}$ is the cage size), 
one separates the total trajectory into caging motions connected by jumps. 
$l_{\mathrm{c}}^2$ is determined as the value of the particle mean square displacement at the crossover from the plateau behavior to the long-time diffusive behavior. 
In calculating the mean square displacement for 2D samples, the displacement of a particle should be corrected with reference to the coordinates of its cage as \cite{szamei1}: 
\begin{eqnarray}
	\delta\boldsymbol{r}_i^{\mathrm{CR}}(t)&=&[\boldsymbol{r}_i(t)-\boldsymbol{r}_i(0)] \nonumber \\
	&-&\frac{1}{N_{\mathrm{nn}}(i)}\sum_{j=1}^{N_{\mathrm{nn}}(i)}[\boldsymbol{r}_j(t)-\boldsymbol{r}_j(0)] \label{eq12} 
\end{eqnarray}
\begin{equation}
	\langle\delta r^2(t)\rangle^{\mathrm{CR}}=\frac{1}{N}\Sigma_{i=1}^N\left\langle\left[\delta r_i^{\mathrm{CR}}(t)\right]^2\right\rangle \label{eq13}
\end{equation}
where $\boldsymbol{r}_i$ is the position of particle $i$, $N_{\mathrm{nn}}(i)$ is the number of neighbors of particle $i$. 
Here we select particles within the distance at which the pair distribution function $g(r)$ reaches the first minimum as neighbors. 
$l_{\mathrm{c}}^2$ is found to be $0.05$ for the BD conditions, $0.057$ for the 3D MD conditions and $0.048$ for the 2D MD conditions. 
Figure~\ref{fig:8} illustrates the separation of the particle trajectory by cage jumps at the BD condition of $\phi=45\%$ and $\dot{\gamma}=0$.

The spatiotemporal distribution of cage jumps shows remarkable heterogeneity. 
We choose $t_{\chi}$, the time at which the dynamic susceptibility $\chi_4(t)$ \cite{glotzer1} reaches maximum, as the characteristic time to observe clustering of cage jumps. $\chi_4(t)$ is defined as:
\begin{equation}
	\chi_4(t)=\frac{\beta V}{N^2}[\langle Q(t)^2\rangle-\langle Q(t)\rangle^2] \label{eq14}
\end{equation}
where $\beta=1/k_{\mathrm{B}}T$, $Q(t)=\sum_{i=1}^N\sum_{j=1}^Nw(\left|\boldsymbol{r}_i(t)-\boldsymbol{r}_j(0)\right|)$, $w(r)$ is an ``overlap'' function which is unity for $r\leq a$ and zero otherwise. 
The value of the parameter $a$ is set to $0.2$ for the BD system, $0.25$ for the 3D MD system and $0.3$ for the 2D MD system. 
It is important to notice that the overlap function under steady shear needs to be adjusted to 
$w\left(\left|\boldsymbol{r}_i(t)-\boldsymbol{r}_j(0)-\dot{\gamma}\int_0^t\mathrm{d}t^{\prime}y_j(t^{\prime})\boldsymbol{e}_x\right|\right)$.

\begin{figure}
    \includegraphics[width=\linewidth]{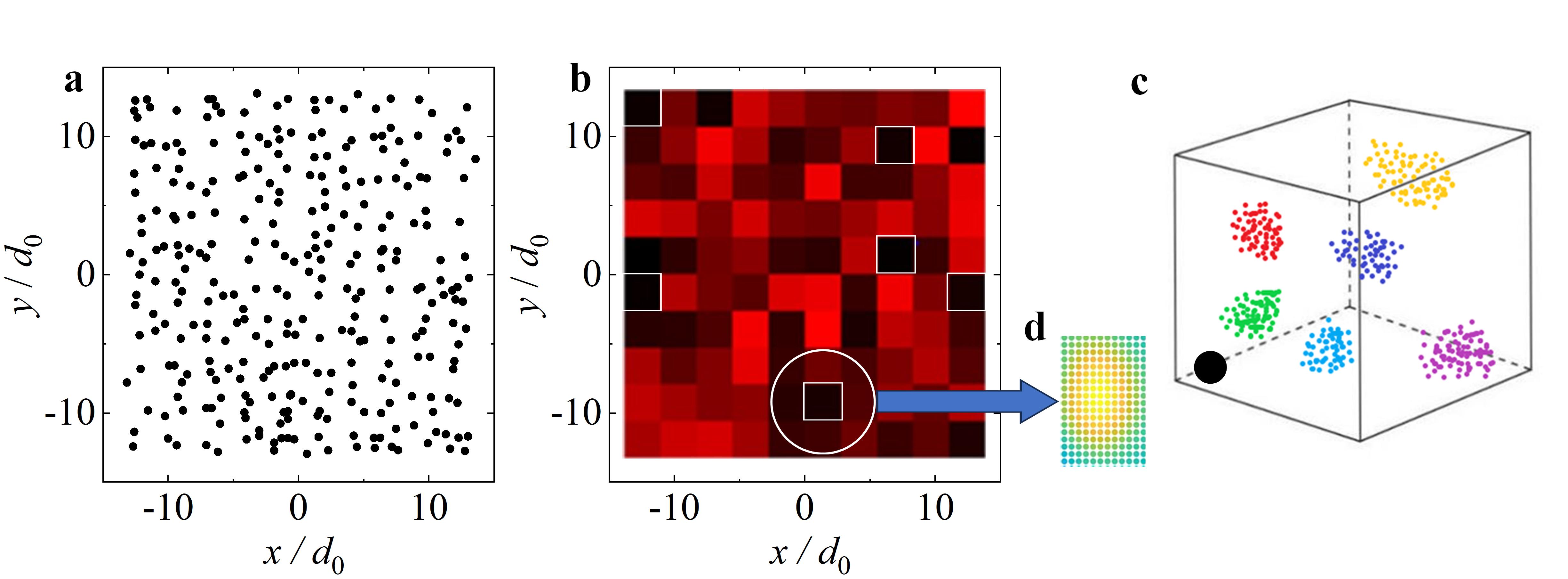}
    \caption{\label{fig:9} 
    \textbf{\textbar Illustration of coarse sieve and undersize sieve.} 
    (\textbf{a}) Spatial distribution of cage jumps during a time interval $t_{\chi}$ inside a slice of thickness $1.5d_0$ at the BD condition of $\phi=45\%$ and $\dot{\gamma}\tau_{0}=3$. 
    (\textbf{b}) Spatial density field of cage jumps coarse-grained from \textbf{a}. 
    Colors of blocks from red to black indicates increase of density. 
    Blocks surrounded by white square frames are those of local density maximum. 
    (\textbf{c}) Clusters obtained after the coarse sieve procedure. 
    The black sphere indicates the size of $V_{\mathrm{stz}}$. 
    (\textbf{d}) Undersize sieve of the cluster highlighted in \textbf{b} by a white circle. 
    Colors of spheres from yellow to blue indicate decrease of density of cage jumps.}
\end{figure}

Considering the remarkable spatial heterogeneity of cage jumps, we propose the following cluster demarcation procedure based on their spatial density distribution. 
The procedure consists of two steps: the coarse sieve and the undersize sieve. Figure~\ref{fig:9} interprets the procedure at the BD condition of $\phi=45\%$ and $\dot{\gamma}\tau_{0}=3$. 
In following paragraphs, we will introduce the two steps of the procedure in detail.

\noindent (I) Coarse sieve:

As for the coarse sieve process, $\rho(\boldsymbol{r})$, the spatial density field of cage jumps, is coarse-grained with the coarse-graining length $d$. 
For BD conditions, we directly divide the box into blocks of length $d$ and count the number of cage jumps within each block. 
$d$ is selected as $L/10$ for $\dot{\gamma}\tau_{0}=1,3,10$ and $L/20$ for $\dot{\gamma}\tau_{0}=20$, where L refers to the box length. 
For MD conditions, the original spatial distribution of cage jumps is convolved by the coarse-graining function $\phi(r)\propto e^{-r/d}$ to obtain the coarse-grained field $\rho(\boldsymbol{r})$. 
Here, we set $d$ as the value at which the pair distribution function $g(r)$ displays the first minimum. 
With the coarse-grained $\rho(\boldsymbol{r})$, we find the local density maximum $\rho(\boldsymbol{r}_{\max})$, where $\boldsymbol{r}_{\max}$ could be the potential center of one cage-jump cluster. 

To demarcate the cage-jump cluster, it is needed to find a reasonable threshold value of $\rho(\boldsymbol{r})$ representing the background, denoted as $\rho_{\mathrm{th}}$. 
Then, regions with $\rho(\boldsymbol{r})\geq\rho_{\mathrm{th}}$ are possible to be included in cage-jump clusters. 
To find $\rho_{\mathrm{th}}$, we perform a simple iteration: For a given initial time $t_{\mathrm{ini}}$, $\rho_{\mathrm{MAX}}$, the global maximum of $\rho(\boldsymbol{r})$, 
is selected as the initial iteration value $\rho_{0}$. 
Then we calculate $\rho_{\mathrm{avg}}=\mathrm{Avg}(\rho<\rho_{0})$ and assign the value of $\rho_{\mathrm{avg}}$ to $\rho_{1}$, which is the initial value for the next iteration step. 
We repeat this procedure until $|\rho_{\mathrm{avg}}-\rho_{i}|/\rho_{\mathrm{avg}}<C$, where $C$ is an empirical parameter. 
The mean value of all final $\rho_{\mathrm{avg}}$ found with different $t_{\mathrm{ini}}$ is employed as $\rho_{\mathrm{th}}$ for this working condition. 
For the value of $C$, we use $10\%$ for both BD and MD conditions. 
In fact, we have tested different values of $C$ in a range of $5\%$ to $10\%$ for BD simulations, $10\%$ to $20\%$ for 3D MD simulations and $10\%$ to $15\%$ for 2D MD simulations. 
The numerical results of $\xi_{\mathrm{c}}$ and $\eta_{\mathrm{c}}$ only change slightly. 
Moreover, their relevant exponents with respect to $\dot{\gamma}$ remain almost the same.  
 
Now we can determine the coarse-grained boundary of each cage-jump cluster. 
For a given $\rho(\boldsymbol{r})$, we start from the position of a $\boldsymbol{r}_{\mathrm{max}}$, and then explore outwards until the value of $\rho(\boldsymbol{r})$ is lower than $\rho_{\mathrm{th}}$. 
With this procedure, we find the coarse-grained boundary of a cage-jump cluster, within which we have $\rho(\boldsymbol{r})\geq\rho_{\mathrm{th}}$. 

Typically, the yielding of a convective cluster involves several STZs. 
Thus, the size of a cluster cannot be too small. 
We set a lower limit for the volume of the cluster by $V_{\mathrm{c,ll}}=C^{\prime}V_{\mathrm{stz}}$, where $V_{\mathrm{stz}}$ is the characteristic volume of a STZ and $C^{\prime}$ is a numerical factor. 
$V_{\mathrm{stz}}$ is determined by the size of the cage-jump spot (Fig.~\ref{fig:2} f gives several cases of the cage-jump spot). 
We find that the cluster radius $\xi_{\mathrm{c}}$ is not very sensitive to $C^{\prime}$ in certain range of $C^{\prime}$. 
Taking the BD condition of $\phi=45\%$ and $\dot{\gamma}\tau_{0}=3$ as an example, the values of $\xi_{\mathrm{c}}$ are found to be $6.461$, $6.627$ and $6.902$ for $C^{\prime}=2$, $3$, and $4$, respectively. 
In addition, the exponent $\nu_{\mathrm{c}}$ is found to be not sensitive to $C^{\prime}$ either. 
To make the best use of simulation data, we choose $C^{\prime}=2$ for all conditions. 

\noindent (II) Undersize sieve:

This procedure is to refine the coarse-grained boundary found above. 
For a cluster identified by coarse sieve, we divide it into small boxes of length $d^{\prime}$. $d^{\prime}$ is chosen as $L/30$ for 3D systems, and $L/80$ for the 2D system. 
$d^{\prime}$ is close to the corresponding length unit $d_{0}$. 
For each grid point (vertex of small box), we draw a sphere with radius of $\sqrt{2}d^{\prime}/2$, and calculate the density of cage jumps within each sphere. 
Then, a refined cluster boundary can be identified by comparing the local density of cage jumps at each grid point with $\rho_{\mathrm{th}}$.

\subsection{Calculation of PHM}\label{subsec4.3}
To calculate the PHM of a convective cluster, we first separate the cluster from the system at $t=t_0$, where $t_0$ is the time that the cluster starts deforming, by fixing the particles outside the cluster. 
Second, we obtain its inherent structure by the conjugate gradient method with particles outside the cluster fixed. 
Then, we calculate PHMs with reference to the inherent structure according to \cite{lerner1}. 
PHMs are putative displacement fields $\boldsymbol{\pi}$ about a mechanical equilibrium state, for which the cost function:
\begin{equation}
	\mathcal{C}(\mathbf{z})=\frac{(\boldsymbol{\mathcal{H}}:\mathbf{z}\mathbf{z})^2}{\Sigma_{\langle i,j\rangle}\left(\mathbf{z}_{ij}\cdot\mathbf{z}_{ij}\right)^2} \label{eq15} 
\end{equation}
assumes local minima, i.e., they solve
\begin{equation}
	\left.\frac{\partial C}{\partial \mathbf{z}}\right|_{\mathbf{z}=\mathbf{\pi}}=\mathbf{0} \label{eq16}
\end{equation}
where $\mathbf{z}_{ij}\equiv\mathbf{z}_j-\mathbf{z}_i$, the sum in eq.~\ref{eq16} runs over all pairs $\langle i,j\rangle$ of interacting particles. 
$:$ represents double contraction. 
$\boldsymbol{\mathcal{H}}\equiv\partial^{2}\boldsymbol{U}/\partial \boldsymbol{x}\partial \boldsymbol{x}$ is the Hessian matrix. 
PHMs are accessible in any system whose $\boldsymbol{\mathcal{H}}$ is available. 
We solve eq.~\ref{eq16} as follows: Starting with an initial guess $\boldsymbol{\pi}^{(0)}$, repeatedly apply the mapping 
\begin{equation}
	\boldsymbol{\mathcal{F}}(\boldsymbol{\pi})=\frac{\boldsymbol{\mathcal{H}}^{-1}\cdot\boldsymbol{\zeta}(\boldsymbol{\pi})}{\sqrt{\boldsymbol{\zeta}(\boldsymbol{\pi})\cdot\boldsymbol{\mathcal{H}}^{-2}\cdot\boldsymbol{\zeta}(\boldsymbol{\pi})}} \label{eq17}
\end{equation}
where 
\begin{equation}
	\boldsymbol{\zeta}_k(\boldsymbol{\pi})\equiv\sum_{\langle i,j\rangle}(\delta_{jk}-\delta_{ik})(\boldsymbol{\pi}_{ij}\cdot\boldsymbol{\pi}_{ij})\boldsymbol{\pi}_{ij} \label{eq18} 
\end{equation}
until $\boldsymbol{\mathcal{F}}(\boldsymbol{\pi})\simeq\boldsymbol{\pi}$. 
More details about the PHMs are given in \cite{manning1}. 

For all conditions, we count $100$ groups of PHM and cage-jump spots to check if they can well match each other. 
The matching ratio is found to be about $75\%$, suggesting a high correlation between STZs determined by PHM and cage-jump spots.

\begin{acknowledgments}
This research was supported by the National Natural Science Foundation of China (No. 11975136). The Center of High Performance Computing, Tsinghua University is acknowledged for computational resources.
\end{acknowledgments}

\bibliography{ncrev.bib}

\end{document}